\documentclass[nofootinbib]{revtex4}
\usepackage{graphicx}
\usepackage{latexsym}
\def\be{\begin{equation}}
\def\ee{\end{equation}}
\def\bea{\begin{eqnarray}}
\def\eea{\end{eqnarray}}

\begin{document}

\title{Features in Dark Energy Equation of State and Modulations in the Hubble Diagram }

\author{Jun-Qing Xia$^{1}$, Gong-Bo Zhao$^{1}$, Hong Li$^{1}$, Bo Feng$^{2}$
and Xinmin Zhang$^{1}$ }

\affiliation{${}^1$Institute of High Energy Physics, Chinese Academy
of Science, P.O. Box 918-4, Beijing 100049, P. R. China}

\affiliation{ ${}^2$ Research Center for the Early Universe(RESCEU),
Graduate School of Science, The University of Tokyo, Tokyo 113-0033,
Japan}

\date{\today.}

\begin{abstract}

We probe the time dependence of the dark energy equation of state
(EOS) in light of three-year WMAP (WMAP3) and the combination with
other tentative cosmological observations from galaxy clustering
(SDSS) and Type Ia Supernova (SNIa). We mainly focus on cases where
the EOS is oscillating or with local bumps. By performing a global
analysis with the Markov Chain Monte Carlo (MCMC) method, we find
the current observations, in particular the WMAP3 $+$ SDSS data
combination, allow large oscillations of the EOS which can leave
oscillating features on the (residual) Hubble diagram, and such
oscillations are potentially detectable by future observations like
SNAP, or even by the CURRENTLY ONGOING SNIa observations. Local
bumps of dark energy EOS can also leave imprints on CMB, LSS and
SNIa. In cases where the bumps take place at low redshifts and the
effective EOS is close to $-1$, CMB and LSS observations cannot give
stringent constraints on such possibilities. However, geometrical
observations like (future) SNIa can possibly detect such features.
On the other hand when the local bumps take place at higher
redshifts beyond the detectability of SNIa, future precise
observations like Gamma-ray bursts, CMB and LSS may possibly detect
such features. In particular, we find that bump-like dark energy EOS
on high redshifts {\it might} be responsible for the localized
features of WMAP on ranges $l \sim 20-40$, which is interesting and
deserves addressing further.
\end{abstract}

\maketitle

\hskip 1.6cm PACS number(s): 98.80.Es, 98.80.Cq \vskip 0.4cm

\section{Introduction}

The three year Wilkinson Microwave Anisotropy Probe observations
(WMAP3)\cite{Spergel:2006hy,Page:2006hz,Hinshaw:2006,Jarosik:2006,WMAP3IE}
have made so far the most precise probe on the Cosmic Microwave
Background (CMB) Radiations. In the fittings to a constant
equation of state (EOS) of dark energy (DE) $w$, combinations of
WMAP with other cosmological observations are in remarkable
agreement with a cosmological constant (CC) except for the WMAP
$+$ SDSS combination, where $w>-1$ is favored a bit more than
1$\sigma$\cite{Spergel:2006hy}. The measurements of the SDSS power
spectrum\cite{Tegmark:2003uf,sdssfit} in some sense make the most
precise probe of the current linear galaxy matter power spectrum
and will hopefully get significantly improved within the coming
few years. If the preference of $w>-1$ holds on with the
accumulation of cosmological observations this will also help
significantly on our understandings towards dark energy. A
cosmological constant, which is theoretically problematic at
present\cite{SW89,ZWS99}, will NOT be the source driving the
current accelerated expansion and a preferred candidate would be
something like quintessence\cite{quint}. On the other hand, the
observations from the Type Ia Supernova (SNIa) in some sense make
the only direct detection of dark
energy\cite{Riess98,Perl99,Tonry03,Riess04,Riess05,snls} and
currently a combination of WMAP $+$ SNIa or CMB $+$ SNIa $+$ LSS
are well consistent with the cosmological constant and the
preference of a quintessence-like equation state has
disappeared\cite{Spergel:2006hy}. It is noteworthy that in the
combinations with the Lyman $\alpha$ forest
Ref.\cite{Seljak:2006bg} shows that a constant EOS $w<-1$ is
preferred slightly. Moreover when one considers the observational
imprints by dynamical equation of state, an EOS which gets across
$-1$ is mildly favored by the current
observations\cite{Zhao:2006bt,Wang:2006ts}. Intriguingly, we are
also aware that the predictions for the luminosity
distance-redshift relationship from the $\Lambda$CDM model by WMAP
only are in notable discrepancies with the "gold" samples reported
by Riess $et$ $al$\cite{Riess04}. Although the discrepancy might
be due to some systematical uncertainties in the Riess "gold"
sample\cite{Riess04}, this needs to be confronted with the
accumulation of the 5-year SNLS observations\cite{snls} and the
ongoing SNIa projects like the Supernova Cosmology Project (SCP)
and from the Supernova Search Team (SST). Alternatively, this
might be due to the implications of dynamical dark energy with
oscillating equation of state.

Although the temperature-temperature correlation (TT) power of WMAP3
is now cosmic variance limited up to $l \sim 400$  and the third
peak is now detected, the tentative features as discovered by the
first year WMAP\cite{wmap1a,wmap1b,wmap1c} are still present: the
low TT quadrupole and localized oscillating features on TT for $l
\sim 30-50$\cite{Hinshaw:2006}. Although the signatures of glitches
on the first peak as discovered by the first year WMAP have now
become weak, they do exist and go beyond the limited cosmic
variance\cite{Hinshaw:2006}. While for the low WMAP TT quadrupole
many authors are inclined to attribute it to cutoff primordial
spectrum\cite{Contaldi:2003zv,Feng:2003mk}, and even BEFORE the
release of the first year WMAP Ref. \cite{Wang:2002hf} claimed
oscillating primordial spectrum could lead to oscillations around
the first peak of CMB TT power, similar effects {\it might} be due
to features on dark energy rather than inflation. For example,
Ref.\cite{Moroi:2003pq} has attributed the low quadrupole to some
subtle physics of dark energy during inflation.

In the literature there have been many investigations on
inflationary models with broken scale
invariance\cite{KLS85,Starobinsky92,ARS97,LPS98,Chung00,WK00,L00}.
Such features have been invoked to explain the previously observed
feature at $k \sim 0.05$ Mpc$^{-1}$\cite{GSZ00,HHV01,BGSS00,GH01},
or even to solve the small scale problem of the CDM model
\cite{KL00}\footnote{For other solutions to this problem, see e.g.
Refs.~\cite{Spergel PRL}-\cite{Bode APJ} and for a review on this
issue see e.g. Ref.\cite{Tasitsiomi}. }. Moreover
Ref.\cite{Feng:2003nt} has claimed that the pre-WMAP data could not
exclude a large running of the spectral index (for relevant study
see also \cite{Lewis:2002ah}), which has been somewhat dramatically
confirmed by the first year WMAP and WMAP3 in combination with other
observations\cite{Spergel:2006hy,WMAP3IE} except for the case with
the Lyman alpha forest\cite{Viel:2006yh,Seljak:2006bg}. Inflation
and dark energy, both of which describe the accelerated expansion of
the universe, might have some relations and Ref.\cite{PVQinf}
proposed a new picture of quintessential inflation.
Ref.\cite{Xia:2006cr} has made an attempt trying to find such
relations from the observational aspect. While oscillating
primordial spectrum may be responsible for the glitches on CMB,
oscillating EOS of dark energy may be helpful to solve the
coincidence problem of dark
energy\cite{DSPRL,Feng:2004ff,Barenboim:2005np,Barenboim:2004kz}. In
Ref.\cite{Feng:2004ff} in the framework of
Quintom\cite{Feng:2004ad}, an attempt was carried out to unify dark
energy and inflation, meanwhile solving the coincidence problem of
DE.

On theoretical aspect dark energy is among the biggest problem of
modern
cosmology\cite{SW89,ZWS99,Weinberg:1987dv,Dvali:2000hr,Koyama:2005tx,Vilenkin:2006gp,Yokoyama:2001ez}.
Dynamical dark energy models rather than the simple cosmological
constant have attracted more interests in theoretical
studies\cite{Copeland:2006wr}. In cases where the mysterious
component of DE is driven by scalar
fields\cite{quint,kessence,phantom} the EOS is typically not like
that by a cosmological constant, this opens a possibility for us to
tell CC from scalar dark energy models with the cosmological
observations. Moreover the current observations have already opened
a robust window to probe the behavior of dark energy independently,
and in cases when $w\neq -1$ is preferred, dynamical DE models which
satisfy the observations are put
forward\cite{phantom,Feng:2004ad,Guo:2004fq,Li:2005fm,Zhang:2005eg,Zhang:2006ck,relevnt,Copeland:2006wr}.
Given the current ambiguity on theoretical study of dark energy, in
the observational probe of DE one often uses the parametrizations of
EOS.

Previously in the observational probes on oscillating features of
dark energy EOS Ref.\cite{Xia:2004rw} has made some preliminary
fittings to the pre-WMAP3 data and some relevant studies have been
carried out later by Refs.\cite{Barenboim:2004kz,Linder:2005dw}. In
the present paper with the method dealing with the perturbations of
Quintom developed in
Refs.\cite{Zhao:2005vj,Xia:2005ge,Xia:2006cr,Zhao:2006bt}, we aim to
probe the time dependence of the dark energy EOS in light of WMAP3
and the combination with other tentative cosmological observations
from SDSS and SNIa from the Riess "gold" sample or the SNLS
observations. The background evolution and perturbations of Quintom
can be identified with one normal quintessence and one phantom
except for the phantom crossing point, where the natural matching
condition is motivated by the case with a high-dimensional operator
on the kinetic term or the two-field case and the individual sound
speed for each field is assumed to be
unity\cite{Zhao:2005vj,Xia:2005ge,Xia:2006cr,Zhao:2006bt}. In the
present work we mainly focus on cases where the EOS is oscillating
or with local bumps. By performing a global analysis with the Markov
Chain Monte Carlo (MCMC) method, we find the current observations,
in particular the WMAP3 $+$ SDSS data combination, allow large
oscillations of the EOS which can leave oscillating features on the
(residual) Hubble diagram, and such oscillations are potentially
detectable by future observations like SNAP. Local bumps of dark
energy EOS can also leave imprints on CMB, LSS and SNIa. In cases
when the bumps take place at low redshifts and the effective EOS is
close to $-1$, CMB and LSS observations cannot give stringent
constraints on such possibilities. However, geometrical observations
like (future) SNIa can possibly detect such features. On the other
hand when the local bumps take place at higher redshifts beyond the
detectability of SNIa, future precise observations like Gamma-ray
bursts and observations of 21 cm tomography, CMB and LSS may
possibly detect such features. In particular, we find that bump-like
dark energy EOS on high redshifts {\it might} be responsible for the
features of WMAP on ranges $l \sim 20-40$, which is interesting and
deserves addressing further.

The remaining part of our paper is structured as follows: in Section
II we describe the method and the data; in Section III we present
our results on the determination of cosmological parameters with
(WMAP3)\cite{Spergel:2006hy,Page:2006hz,Hinshaw:2006,Jarosik:2006,WMAP3IE},
SNIa \cite{Riess04,snls}, Sloan Digital Sky Survey 3-D power
spectrum (SDSS-P(k)) \cite{Tegmark:2003uf} by global fittings using
the MCMC technique; discussions and conclusions are presented in the
last section.

\section{Method and data}

In the parametrization of oscillating EOS one typically needs four
parameters for the amplitude, center values, phase and the
frequency. And in our analysis we have used
\begin{equation}\label{Wqosc}
w= w_0 + w_1 \sin (w_2 \ln a + w_3)~~.
\end{equation}
The case with $w_0=-1$ and $w_1=0$ corresponds to the cosmological
constant.

The method we adopt is based on the publicly available Markov
Chain Monte Carlo package
\texttt{CosmoMC}\cite{Lewis:2002ah,IEMCMC}, which has been
modified to allow for the inclusion of dark energy perturbations
with EOS getting across $-1$\cite{Zhao:2005vj}. Our most general
parameter space is
\begin{equation}\label{para}
    \textbf{p}\equiv(\omega_{b}, \omega_{c}, \Theta_S, \tau,w_{0}, w_{1}, w_2, w_3, n_{s}, \log[10^{10}
    A_{s}])~,
\end{equation}
where $\omega_{b}=\Omega_{b}h^{2}$ and $\omega_{c}=\Omega_{c}h^{2}$
are the physical baryon and cold dark matter densities relative to
critical density, $\Theta_S$ is the ratio (multiplied by 100) of the
sound horizon to the angular diameter distance at decoupling, $\tau$
is the optical depth, $A_{s}$ is defined as the amplitude of initial
power spectrum and $n_{s}$ measures the spectral index. Assuming a
flat Universe motivated by inflation and basing on the Bayesian
analysis, we vary the above 10 parameters and fit to the
observational data with the MCMC method. We
take the weak priors as: 
$\tau<0.8, 0.5<n_{s}<1.5, -4<w_{0}<1, -10<w_{1}<10, 0<w_2<20,
-\pi/2<w_3<\pi/2 $, a cosmic age tophat prior as 10 Gyr$<t_{0}<$20
Gyr. The choice of priors on $w_0, w_1,w_2,w_3$ have been set to
allow for spread in all of the parameters simultaneously.
Furthermore, we make use of the HST measurement of the Hubble
parameter $H_0 = 100h \quad \text{km s}^{-1} \text{Mpc}^{-1}$
\cite{freedman} by multiplying the likelihood by a Gaussian
likelihood function centered around $h=0.72$ and with a standard
deviation $\sigma = 0.08$. We impose a weak Gaussian prior on the
baryon and density $\Omega_b h^2 = 0.022 \pm 0.002$ (1 $\sigma$)
from Big Bang nucleosynthesis\cite{bbn}. The bias factor of LSS has
been used as a continuous parameter to give the minimum $\chi^2$.

In our calculations we have taken the total likelihood to be the
products of the separate likelihoods of CMB, SNIa and LSS.
Alternatively defining $\chi^2 = -2 \log {\bf \cal{L}}$, we get \be
\chi^2_{total} = \chi^2_{CMB}+ \chi^2_{SNIa}+\chi^2_{LSS}~~~~ .\ee
In the computation of CMB we have included the three-year WMAP
(WMAP3) data with the routine for computing the likelihood supplied
by the WMAP team \cite{WMAP3IE}. To be conservative but more robust,
in the fittings to the 3D power spectrum of galaxies from the
SDSS\cite{Tegmark:2003uf} we have used the first 14 bins only, which
are supposed to be well within the linear regime\cite{sdssfit}. In
the calculation of the likelihood from SNIa we have marginalized
over the nuisance parameter\cite{DiPietro:2002cz}. The supernova
data we use are the "gold" set of 157 SNIa published by Riess $et$
$al$ in \cite{Riess04} and the 71 high redshift type Ia supernova
discovered during the first year of the 5-year Supernova Legacy
Survey (SNLS)\cite{snls} respectively. In the fittings to SNLS we
have used the additional 44 nearby SNIa, as also adopted by the SNLS
group\cite{snls}. Also to be conservative but more robust, we did
not try to combine SNLS with the Riess sample simultaneously for
cosmological parameter constraints, namely in one case for SNIa
fitting we use SNLS data only and in another case the Riess sample
only. For each regular calculation, we run 6 independent chains
comprising of 150,000-300,000 chain elements and spend thousands of
CPU hours to calculate on a cluster. The average acceptance rate is
about 40\%. And for the convergence test typically we get the chains
satisfy the Gelman and Rubin\cite{GR92} criteria where R-1$<$0.1.

In our study for future perspectives on features of dark energy we
have used the cosmic-variance\cite{Bond:1987ub} limited CMB TT
spectrum up to $l = 2000$. For SNIa we have used SNAP\cite{snap}
simulations\footnote{ SNAP is one of the several candidate mission
concepts for the Joint Dark Energy Mission (JDEM). Nowadays there
have been many proposed dark energy surveys\cite{JDEM}.} and for
LSS, we have adopted the LAMOST\cite{lamost} simulations. In the
remaining part of this paper the fiducial power law $\Lambda$CDM
model adopted is as follows:
\begin{equation}\label{fid}
  (\omega_{b}, \omega_{c}, h, z_r, n_{s},
  A_{s}) =(0.022,0.12, 0.7, 12, 1, 2.3\times 10^{-9}  )~~,
\end{equation}
where $z_r$ is the reionization redshift and the slightly different
notations from previous Eq. (\ref{para}) are due to the difference
in the CAMB\cite{Lewis:1999bs,IEcamb} and
\texttt{CosmoMC}\cite{Lewis:2002ah,IEMCMC} default parameters. Such
a fiducial model will be used to generate future CMB, SNIa and LSS
data. In addition the illustrative figures will also be generated
with such background parameters. Moreover in generating the
illustrative figures on linear power spectrum of LSS we have fixed
the bias factor to be unity.

 The projected satellite SNAP (Supernova / Acceleration
Probe) would be a space based telescope with a one square degree
field of view with 1 billion pixels. It aims to increase the
discovery rate for SNIa to about 2,000 per year\cite{snap}. The
simulated SNIa data distribution is taken from Refs.
\cite{kim,tao,Li:2005zd}. As for the error, we follow the ref.
\cite{kim} which takes the magnitude dispersion $0.15$ and the
systematic error $\sigma_{sys}(z)=0.02\times z/1.7$, and the whole
error for each data is
\begin{equation}
\sigma_{mag}(z_i)=\sqrt{\sigma_{sys}^2(z_i)+\frac{0.15^2}{n_i}}~~,
\end{equation}
where $n_i$ is the number of supernova in the i'th redshift bin.

The Large Sky Area Multi-Object Fiber Spectroscopic Telescope
(LAMOST) project as one of the National Major Scientific Projects
undertaken by the Chinese Academy of Science, aims to measure
$\sim 10^7$ galaxies with mean redshift $z \sim 0.2$\cite{lamost}.
In the measurements of large scale matter power spectrum of
galaxies there are generally two statistical errors: sample
variance and shot noise. The uncertainty due to statistical
effects, averaged over a radical bin $\Delta k$ in Fourier space,
is \cite{9304022}
\begin{equation}
\label{eqn:dPK} (\frac{\sigma_P}{P})^2  = 2\times \frac{(2
\pi)^3}{V}\times \frac{1}{4 \pi k^2 \Delta k}\times (1+
\frac{1}{\bar{n}P})^2~~.
\end{equation}
The initial factor of 2 is due to the real property of the density
field, $V$ is the survey volume and $\bar{n}$ is the mean galaxy
density. In our simulations for simplicity and to be conservative,
we use only the linear matter power spectrum up to $k \sim 0.15$
$h$ Mpc$^{-1}$.

For the cases with future cosmic variance limited CMB and LAMOST, we
only show the error bars for illustrations, and a further analysis
with fittings is not the aim of the present paper. For the studies
on bump-like dark energy EOS, we also plot the illustrative results
rather than make global fittings. And the parametrized EOS takes the
following form:
\begin{equation}\label{Wqbump}
w= w_0 + A (\ln a - \lambda)^3 \exp(- (\ln a - \lambda)^4/d)~~.
\end{equation}
We should point out here that the parametrization in
Eq.(\ref{Wqbump}) is a specific example only and there are certainly
different parametrizations to illustrate the bump-like features in
dark energy EOS and the resulting cosmological imprints might be
different.

\section{Results }

\begin{figure}
\includegraphics[scale=0.5]{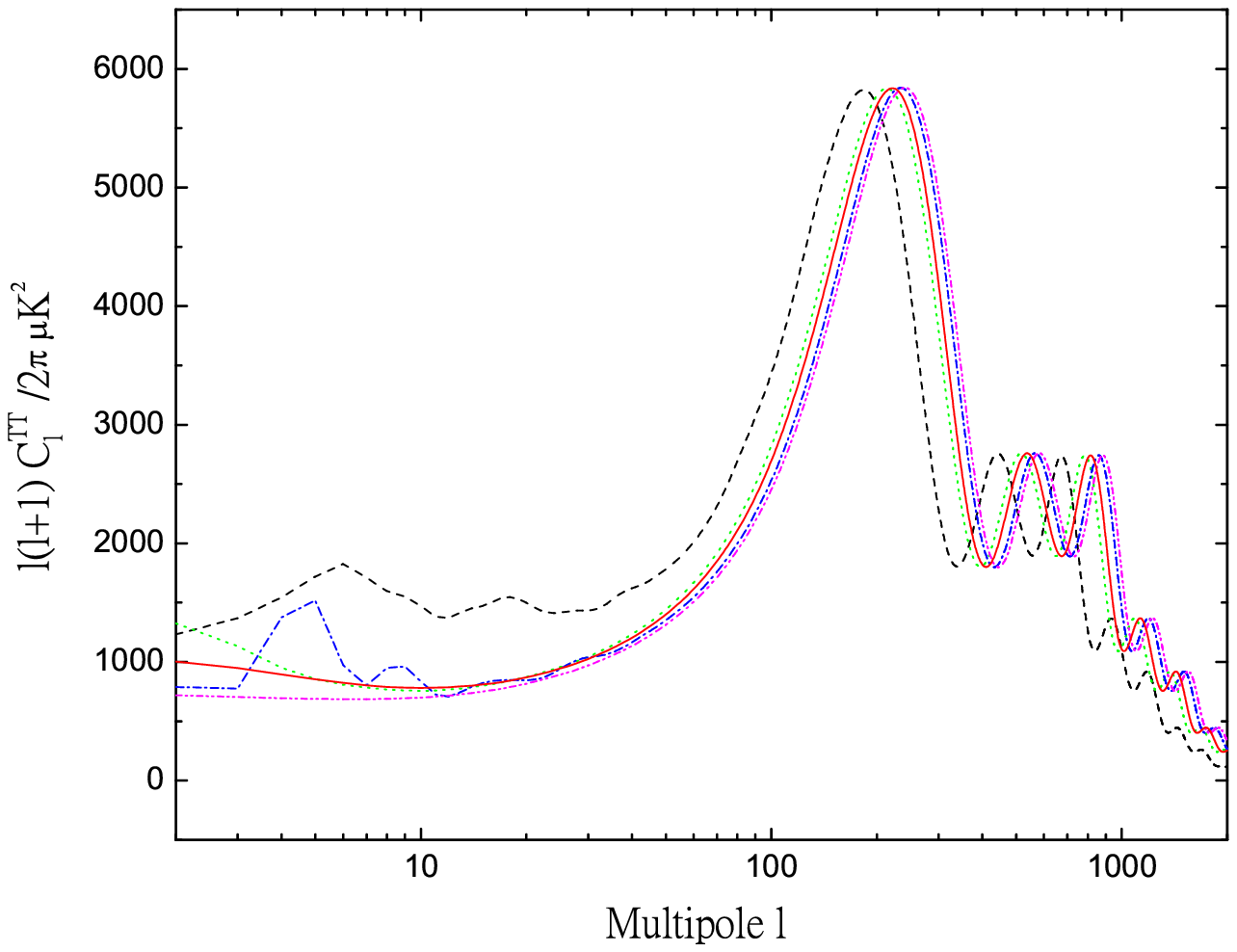}
\includegraphics[scale=0.5]{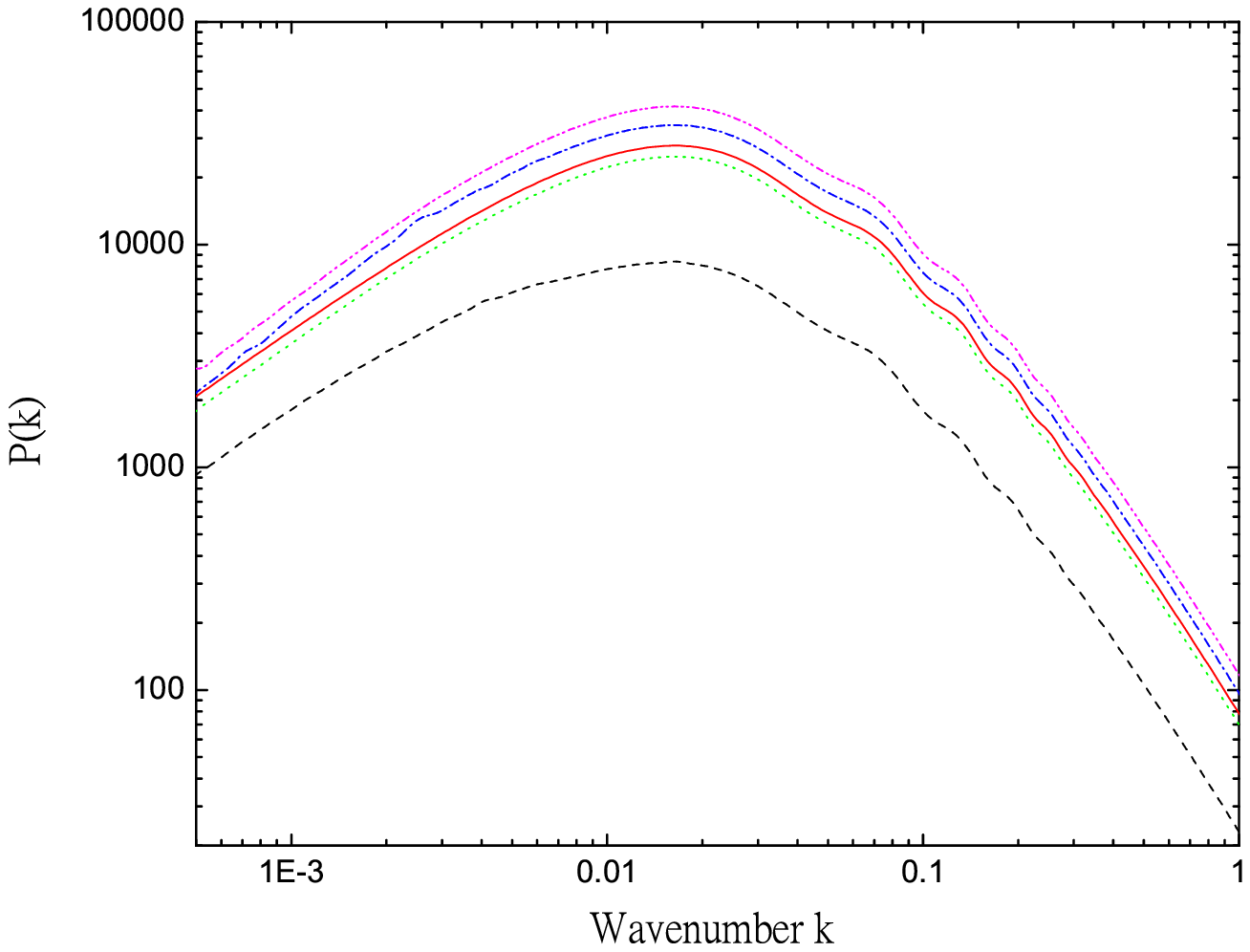}
\includegraphics[scale=0.5]{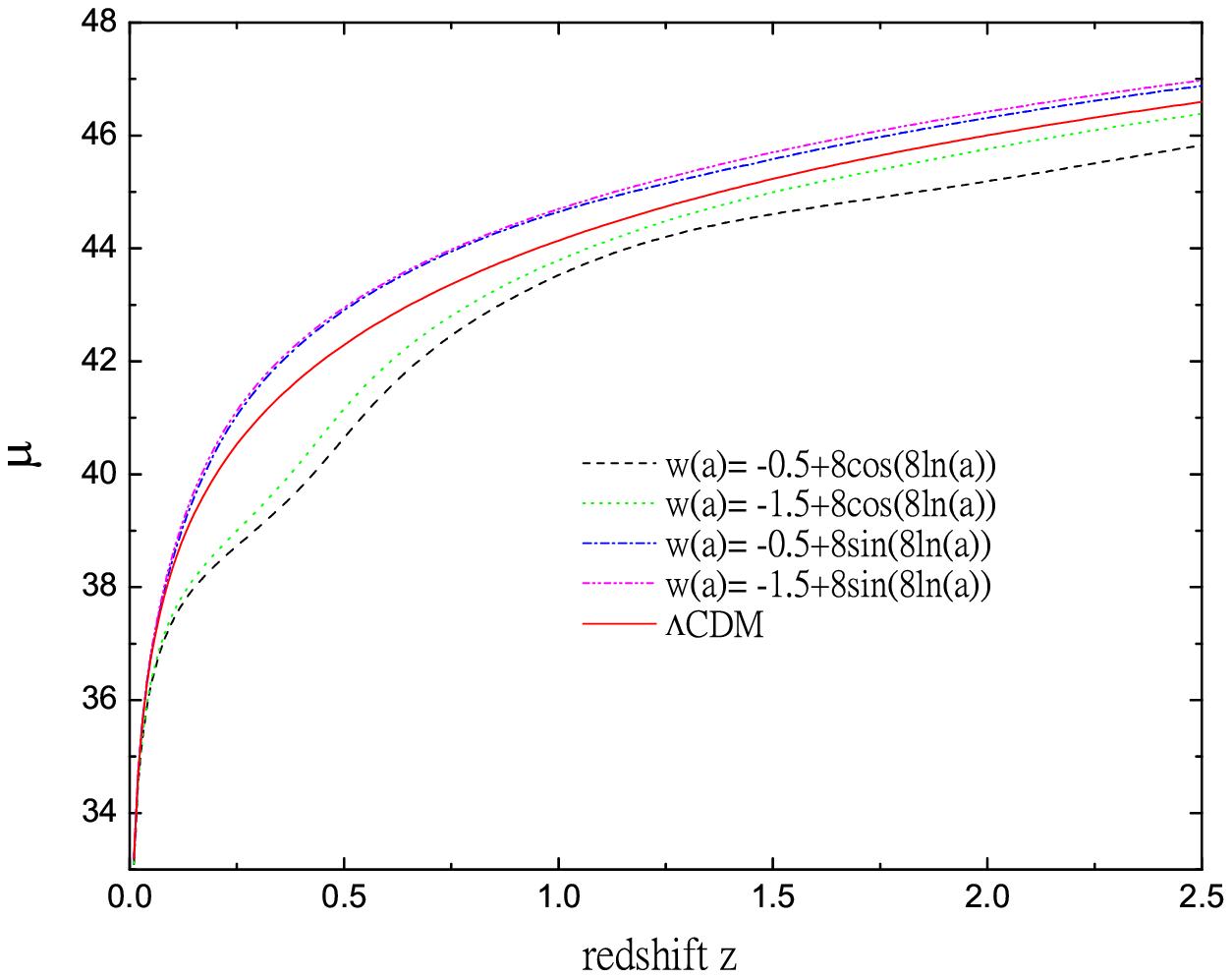}
\caption{Illustrative imprints of oscillating dark energy equation
of state (EOS) on CMB (top left), LSS (top right)
 and on the Hubble diagram (lower panel) \label{fig:il} .
 }
\end{figure}

We start with the oscillating case. First of all in
Fig.\ref{fig:il} we delineate the illustrative imprints of
oscillating dark energy equation of state (EOS) on CMB (top left),
LSS (top right)
 and on the Hubble diagram (lower panel).In the Hubble diagram the distance modulus
$\mu$ is defined as the apparent magnitude $m$ minus the absolute
magnitude $M$: \begin{equation} \mu \equiv m-M=5\lg \left(
\frac{d_{L}}{1 Mpc} \right) + 25,
\end{equation}
with $d_L$ being the luminosity distance:
\begin{equation}
\frac{d_{L}}{1+z} = \int^{z}_{0}\frac{dz'}{H(z')} .
\end{equation}
In comparison the imprints by the $\Lambda$CDM cosmology have also
been displayed.
 The main contribution of dark energy on CMB is on the geometrical
 angular diameter distance to the last scattering surface. This in
 turn determines the locations of CMB peaks. In some
cases the effects on the large scale matter power spectrum of LSS
are also significant. In Fig.\ref{fig:il} we can find for cases
where $w_0=-0.5$, the effects on CMB turn out to be most eminently
modulated. This is mainly due to the fact that in such cases as
$w_0$ deviates significantly from $-1$ and $w$ has been {\it
relatively} matter-like, the contributions to large scale CMB are
significant due to the Integrated Sachs-Wolfe (ISW) effects.  An
oscillating EOS of dark energy can leave somewhat similar imprints
as oscillating primordial spectrum, as explored in Ref.
\cite{Wang:2002hf}, although the difference is also noteworthy. We
find interestingly that in cases $w_0=-1.5$, the effects are not
as large as $w_0=-0.5$, which is in part due to the effects of
dark energy perturbations and consistent with the previous
analysis in Ref. \cite{Zhao:2005vj}. The effects of dark energy on
CMB (and LSS) are mainly geometrical effects and can sometimes be
understood from the formula on $w_{eff}$\cite{Wang:1999fa}:
\begin{equation}\label{weff}
    w_{eff}\equiv\frac{\int da \Omega(a) w(a)}{\int da
    \Omega(a)}~~.
\end{equation}
On the other hand the Hubble diagram being displayed with the
redshift, the effects of  dark energy, in cases when DE dominates
the Universe the EOS has oscillating features, can leave
oscillations on the the Hubble diagram.

For the next step we show the results from our global fittings on
the oscillating EOS in Eq.(\ref{Wqosc}). In Table 1 we delineate the
mean $1\sigma$ constrains on the relevant cosmological parameters
using different combination of WMAP, SNIa and SDSS. Shown together
are those with maximum likelihood (dubbed ML) and those which give
the most eminent oscillating effects in the 2$\sigma$ allowed
regions (dubbed "Most" in the table). For simplicity the parameters
related to bayon fractions and the primordial spectrum are not
displayed, and the shown parameters are enough for the study on the
Hubble diagrams below. Typically in the realizations of MCMC as the
cosmological parameters are not exactly gaussian distributed, the
center mean values are different from the best fit
cases\footnote{More detailed discussions are available at
http://cosmocoffee.info/  .}. We find that although the best fit
cases are given by oscillating EOS, a cosmological constant (
$w_0=-1,w_1=0$ ) is well within 1$\sigma$ for all the three data
combinations. The accumulation of the observational data will help
to break such a degeneracy. It is very interesting that $w_1$ is
much better constrained by SNLS (together with the 44 low redshift
SNIa) rather than by the Riess "gold" sample. On the other hand the
background parameter $H_0$ is better constrained by the Riess "gold"
sample. One can also find that the frequency of oscillations, $w_2$,
is relatively the worst constrained parameter. We will turn to this
in more details in the remaining part of the present paper.

\begin{table*}
TABLE 1. Mean $1\sigma$ constrains on cosmological parameters using
different combination of WMAP, SNIa and SDSS. Shown together are
those with maximum likelihood (ML) and those which give the most
eminent oscillating effects within the 2$\sigma$ allowed regions
(Most).

\begin{center}
\begin{tabular}{|c|ccc|ccc|ccc|}

  \hline
&\multicolumn{3}{c|}{ WMAP+SDSS }&\multicolumn{3}{c|}{ WMAP+SDSS+Riess }&\multicolumn{3}{c|}{ WMAP+SDSS+SNLS }\\
&Mean&ML&Most&Mean&ML&Most&Mean&ML&Most\\

\hline

$w_0$ &$-0.805^{+0.413}_{-0.394}$ &$-0.465$&$-0.250$&$-0.845^{+0.308}_{-0.222}$ &$-0.508$&$-0.450$&$-0.886^{+0.283}_{-0.215}$&$-0.485$&$-0.400$ \\

$w_1$     &$0.874^{+2.910}_{-3.068}$ &2.62&5.80&$0.609^{+1.228}_{-1.580}$ &1.58&2.90&$0.303^{+0.801}_{-0.792}$&1.29&1.90 \\

$w_2$ &$11.6\pm5.9$ &18.3&11.5   &$10.8^{+5.3}_{-6.0}$&12.0&10.8 &$9.52^{+10.47}_{-9.49}$ &8.28&9.50  \\

$w_3$  &$0.179^{+0.990}_{-1.730}$&$-0.0707$&1.5000 &$-0.0446^{+0.8646}_{-1.5203}$ &$-0.131$&1.400&$0.0831^{+0.9316}_{-1.6505}$ &0.356&1.400\\

$\Omega_m$ &$0.345^{+0.060}_{-0.057}$ &0.332&0.350&$0.308^{+0.033}_{-0.032}$ &0.325&0.300&$0.288^{+0.033}_{-0.032}$&0.312&0.300\\

$\Omega_{\Lambda}$ &$0.655^{+0.057}_{-0.060}$ &0.668&0.650
&$0.692^{+0.032}_{-0.033}$ &0.675&0.700&$0.712^{+0.032}_{-0.033}$&0.688&0.700\\

$H_0$  &$61.8^{+5.6}_{-5.9}$&62.0&62.0 &$65.6\pm3.5$&63.1&65.5 &$68.2^{+3.6}_{-3.7}$&62.5&68.0\\

  \hline
\end{tabular}
\end{center}
\end{table*}

\begin{figure}[htbp]
\begin{center}
\includegraphics[scale=0.5]{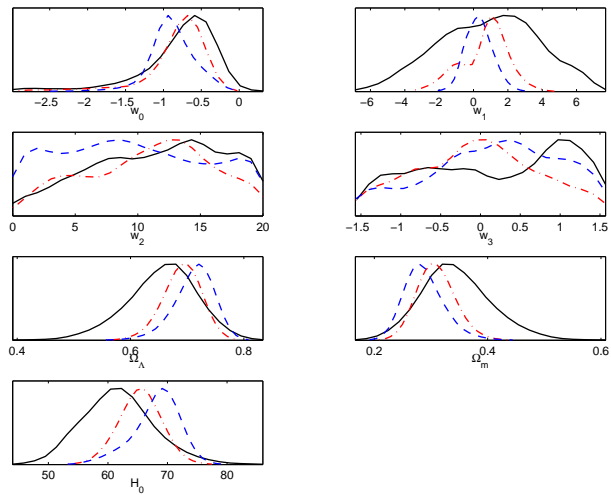} \caption{ One dimensional posterior constraints on the
parametrized EOS: $w= w_0 + w_1 \sin (w_2 \ln a + w_3)$ and on the
relevant background cosmological parameters. The black solid lines
are constraints from combined analysis of WMAP3 $+$ SDSS. The red
dash-dot lines are from WMAP3 $+$ SDSS $+$ Riess sample and the
blue dashed lines are from WMAP3 $+$ SDSS $+$ SNLS combined
analysis \label{fig:1dosc} . }
\end{center}
\end{figure}

In Fig.\ref{fig:1dosc} we delineate the one dimensional posterior
constraints on the oscillating EOS in Eq.(\ref{Wqosc}), showing
together some relevant background cosmological parameters. The black
lines are constraints from combined analysis of WMAP3 $+$ SDSS. The
red lines are from WMAP3 $+$ SDSS $+$ Riess sample and the blue are
from WMAP3 $+$ SDSS $+$ SNLS combined analysis. We can find that in
some sense  $w_2$ and $w_3$ are not well constrained by the current
observations. As $w_3$ represents the phase of oscillations and the
range ( $-\pi/2, \pi/2$ ) is the largest 1-period limit (note we
have allowed $w_1$ to be both positive and negative), our prior on
$w_2$, though as large as given above, turn out to be somewhat too
optimistic.

\begin{figure}
\includegraphics[scale=0.6]{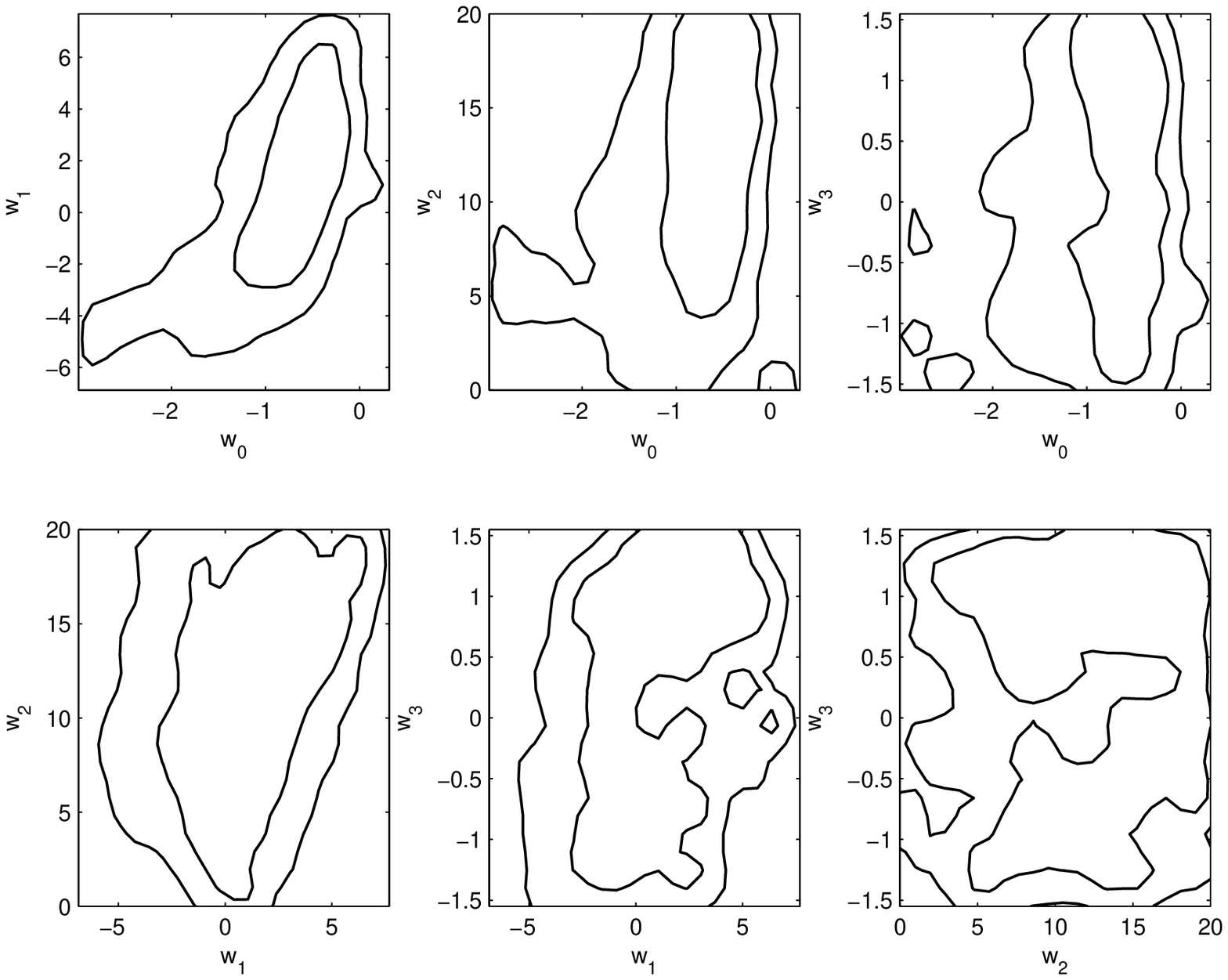}
\includegraphics[scale=0.45]{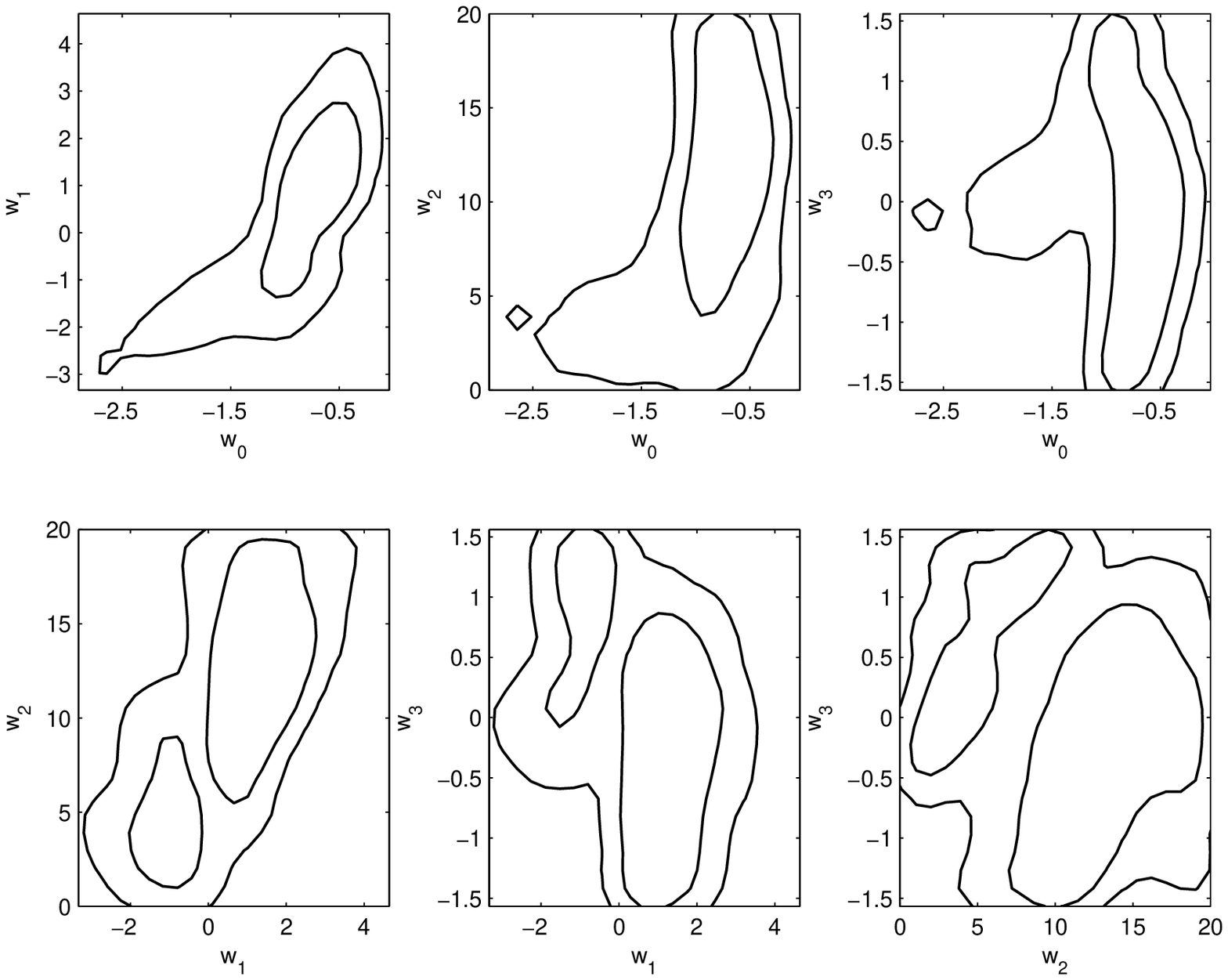}
\includegraphics[scale=0.45]{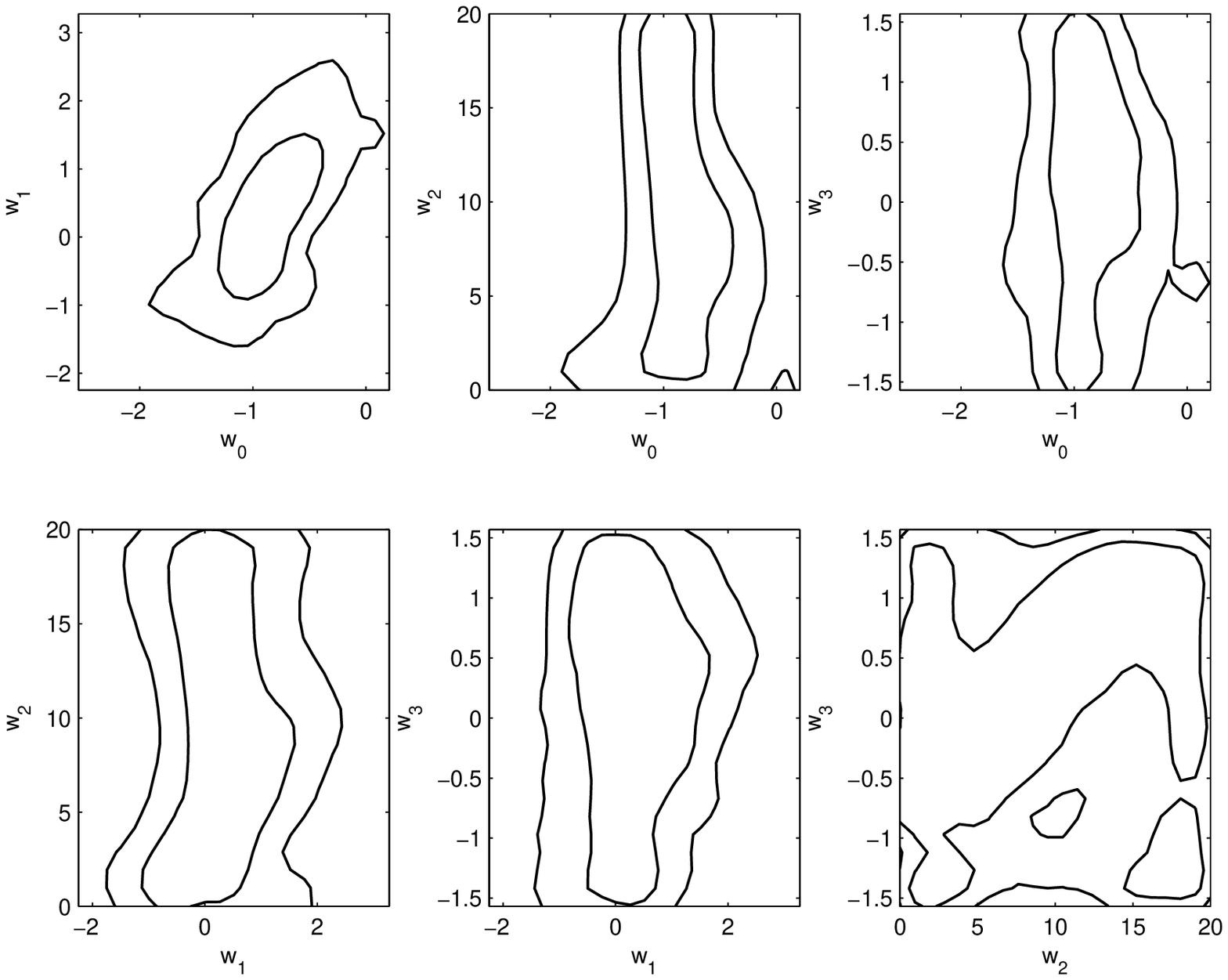}
\caption{ Two dimensional posterior constraints on the parametrized
EOS: $w= w_0 + w_1 \sin (w_2 \ln a + w_3)$ at 68\% and 95\% C.L. The
upper panel is on the constraints from combined analysis of WMAP3
$+$ SDSS. The lower left panel is from WMAP3 $+$ SDSS $+$ Riess
sample and the lower right from WMAP3 $+$ SDSS $+$ SNLS combined
analysis \label{fig:2dosc} .
 }
\end{figure}

The effects are more eminent in the resulting two dimensional
contours. In Fig.\ref{fig:2dosc} we plot the corresponding two
dimensional posterior constraints.  The upper panel is on the
constraints from the combined analysis of WMAP3 $+$ SDSS. The
lower left panel is from WMAP3 $+$ SDSS $+$ Riess sample and the
lower right
from WMAP3 $+$ SDSS $+$ SNLS combined analysis. 

\begin{figure}[htbp]
\begin{center}
\includegraphics[scale=1]{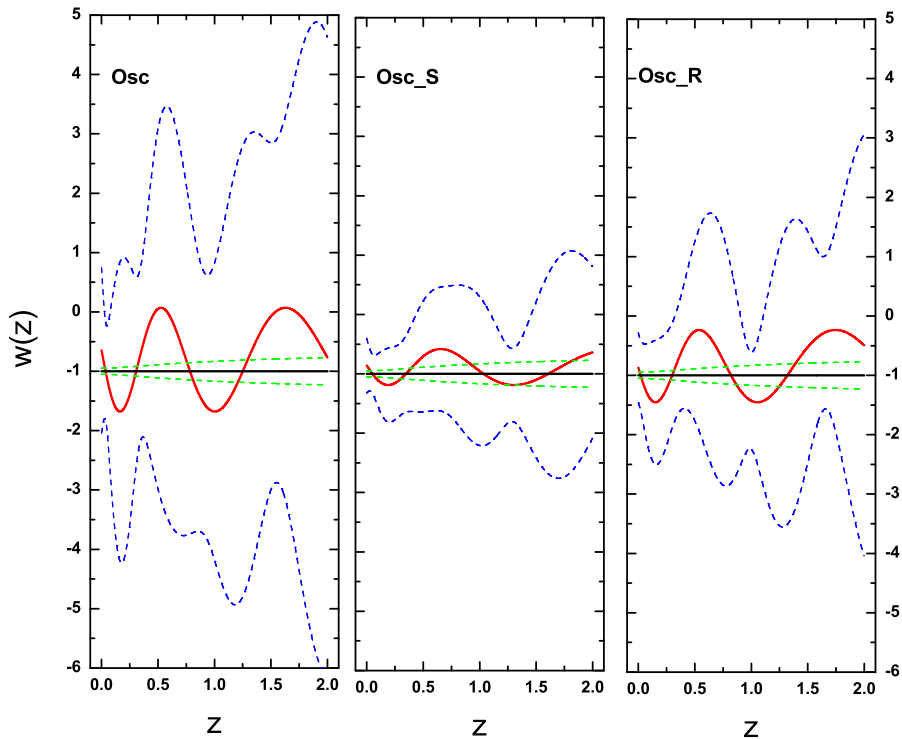}
\caption{ Resulting posterior constraints on the low-redshift
behavior of the parametrized EOS: $w= w_0 + w_1 \sin (w_2 \ln a +
w_3)$. The red lines are given by the mean center values as shown
in Table 1 and the outside blue dashed lines are the 1$\sigma$
allowed regions by WMAP3 $+$ SDSS (left), WMAP3 $+$ SDSS $+$ SNLS
(center) and WMAP3 $+$ SDSS $+$ Riess sample (right). The inside
green dashed lines are the illustrative 1$\sigma$ explored regions
by future SNAP \label{fig:wzx} .}
\end{center}
\end{figure}

It is understandable that in cases one or two parameters are not
well constrained by the observations, changing the priors (in our
case on $w_2$) will inevitably affect the final results. Such a
feature has also appeared in previous investigations on the
oscillating primordial spectrum, regardless of fitting methods one
uses such as grids\cite{Elgaroy:2003gq} or
MCMC\cite{Okamoto:2003wk,Easther:2004vq}. For the grid case the
problem is how small one should set the minimum grid, and there will
be finer features on scales smaller than the minimum
steps\footnote{For the grid case in the fittings to give the first
year WMAP preferred running of the spectral index, in Ref.
\cite{Feng:2003mk} in the case with an oscillating inflaton
potential, the fractor-like structure is also present. Similar
potentials can also lead to oscillations on the primordial spectrum,
as previously investigated in Ref.\cite{Wang:2002hf}. And it is
easily understood that in cases where the oscillations of the
primordial spectrum have a large period, the resulting effects on
CMB and LSS relevant scales will be running-like. Note besides
inflation similar behavior also applies to the case of dark energy.
}. For our conventional cases with
MCMC\cite{Xia:2005ge,Xia:2006cr,Zhao:2006bt,Feng:2006dp} to get the
converged results, we test the convergence of the chains by Gelman
and Rubin\cite{GR92} criteria and typically get R-1 to be of order
0.01, which is more conservative than the recommended value
R-1$<$0.1. However in the present case with an oscillating EOS, in
the combinations with SNLS the maximum value of  R-1 is 0.04, for
the combination with  Riess "gold" sample, we have value: R-1 $\sim$
0.1 and for the the case with WMAP3 $+$ SDSS we have the maximum R-1
to be 0.25 instead, although for the last case we have run much
longer time for the chains. This also shows that for the case with
oscillations at least for WMAP3 $+$ SDSS our chains are not well
converged. On the other hand this shows that for the current
observations a large oscillation on the dark energy EOS is allowed.
In Ref.\cite{Okamoto:2003wk} the authors quoted a value where R-1
$<0.1$ (their R-1 is slightly different from ours) and the
structures of our resulting two-dimensional figures are similar to
those by Refs.\cite{Okamoto:2003wk,Easther:2004vq}. These have
implied that in the presence of oscillations typically the current
data are not good enough to well break the degeneracy among the
parameters. In our case although the parameters like $n_s,$ $w_0$,
$\Omega_{DE}$, $\Omega_m $ and $H_0$ have already been well
constrained, R-1 is in some cases relatively large due to the
parameters $w_2$ and $w_3$. This has in turn led to the separate
likelihood spaces in the three data combinations at 68\% and 95\%
C.L. Interestingly in the WMAP3 $+$ SDSS $+$ Riess sample the 1
$\sigma$ contours clearly separate into two different peaks in the
lower three panels, resembling the contours in Ref.
\cite{Okamoto:2003wk}. Such a behavior is less eminent in the WMAP3
$+$ SDSS $+$ SNLS sample, which nevertheless does exist in the $w_2
- w_3$ contour. We should point out that given the priors and the
R-1 value specified above, our results are robust. In
Fig.\ref{fig:wzx} we delineate the resulting posterior $1 \sigma$
constraints on the low-redshift behavior of the oscillating EOS,
with the three different data combinations. The red lines are given
by the mean center values as shown in Table 1 and the blue dashes
lines are the $1\sigma$ allowed regions. The green dashed lines are
the illustrative $1 \sigma$ explored regions by future
SNAP\cite{snap}. Thus future SNIa observations like SNAP can help
significantly to break the degeneracy, and in some sense detect such
oscillating features of dark energy EOS.

\begin{figure}
\includegraphics[scale=0.6]{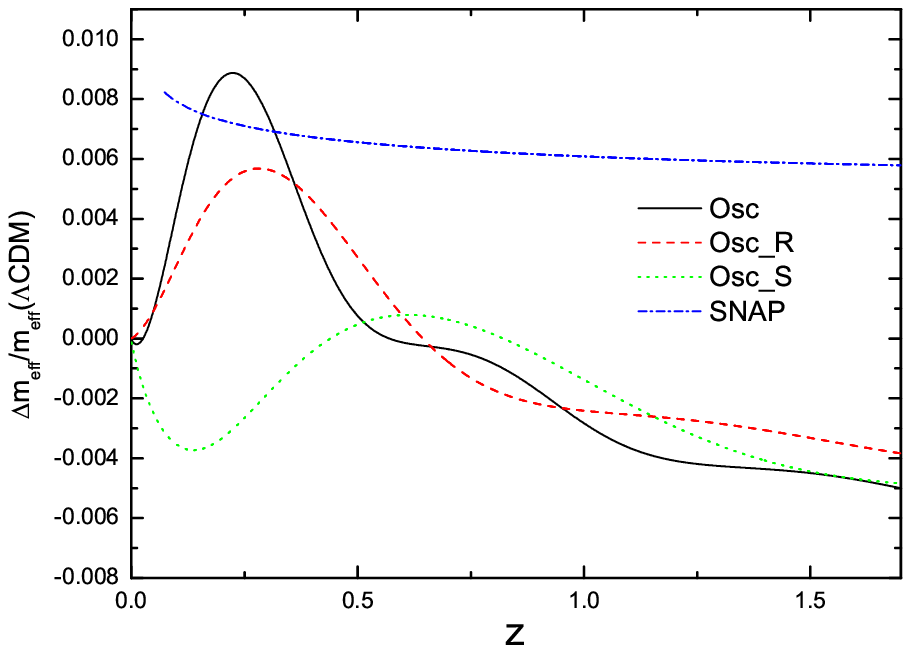}
\includegraphics[scale=0.6]{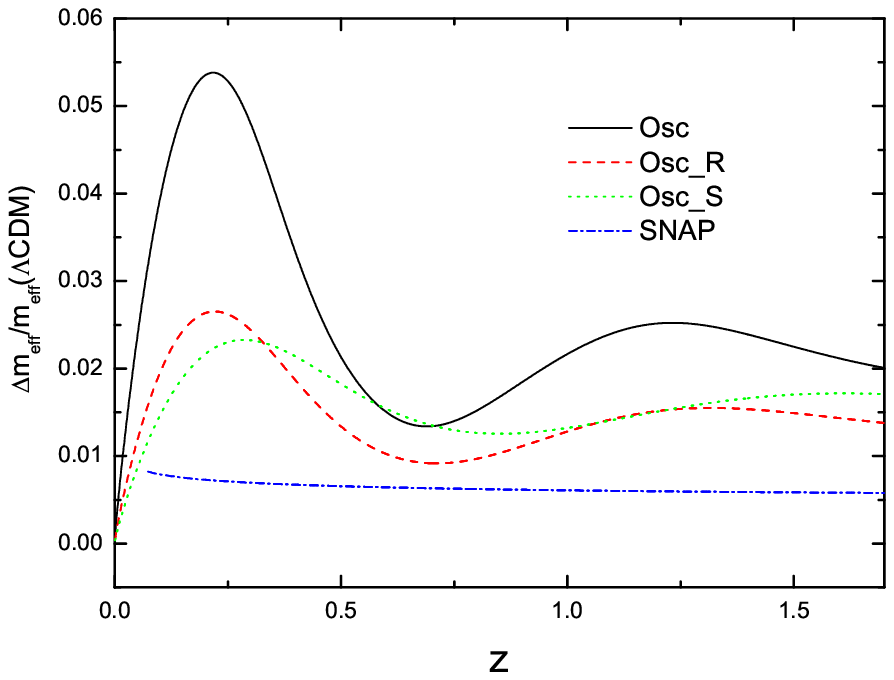}
\includegraphics[scale=0.6]{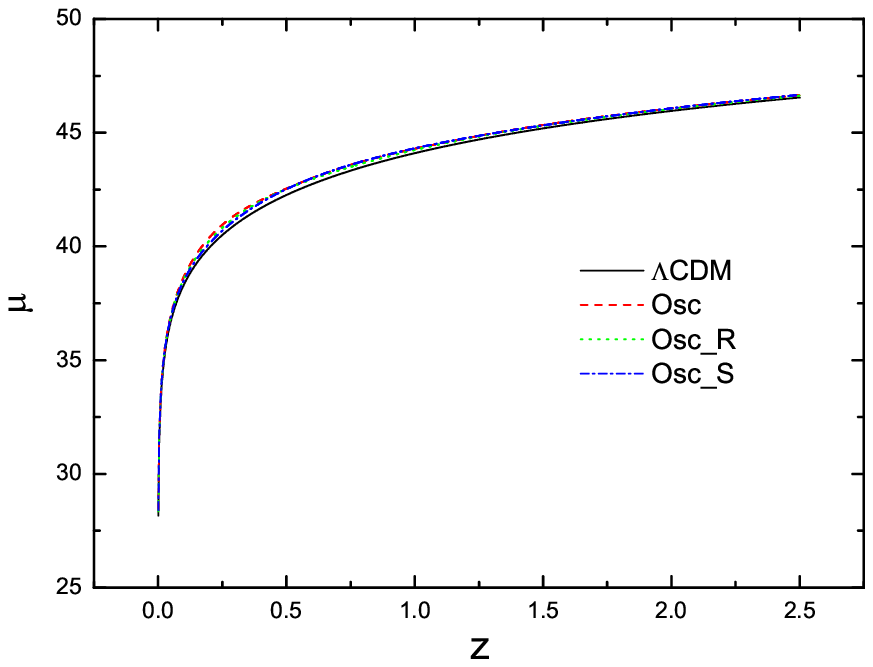}
\includegraphics[scale=0.6]{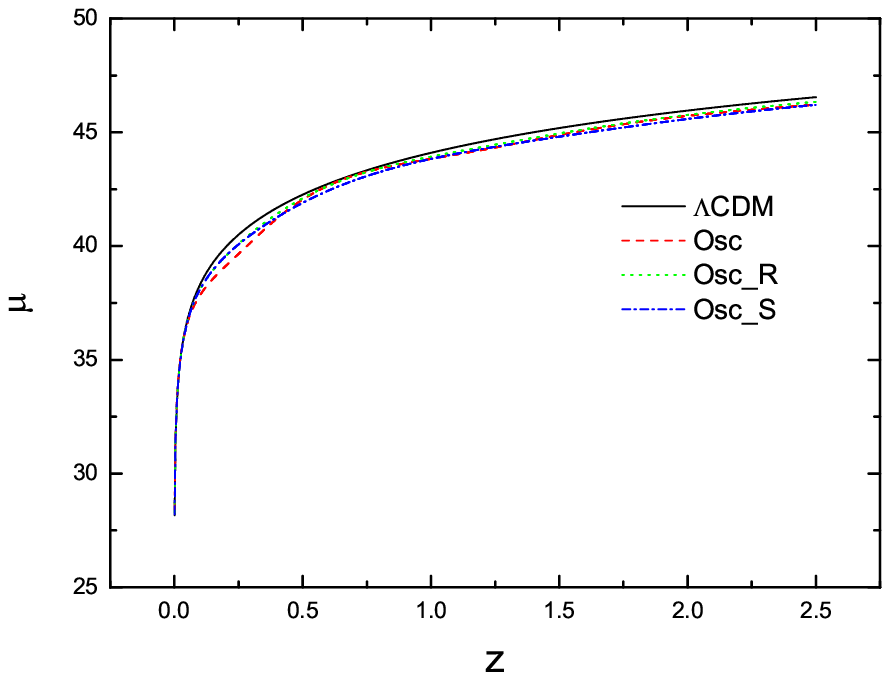}
\caption{ Resulting imprints of the parametrized EOS: $w= w_0 +
w_1 \sin (w_2 \ln a + w_3) $ on the residue Hubble diagram (upper
panels) and the Hubble diagram (lower panels). In the left panels,
the lines dubbed "Osc" are given by the best fit values WMAP3 $+$
SDSS, the underlined "S" is by WMAP3 $+$ SDSS $+$ SNLS and "R" by
WMAP3 $+$ SDSS $+$ Riess sample. The right panels show the
corresponding cases where within the 2$\sigma$ allowed regions the
oscillating effects are relatively eminent. In the upper panels
the blue lines dubbed "SNAP" illustrate the detectability of
future SNAP \label{fig:Hubble} .
 }
\end{figure}

In Fig.\ref{fig:Hubble} we delineate the corresponding imprints on
the residue Hubble diagram (upper panels) and the Hubble diagram
(lower panels). In the left panels, the lines dubbed "Osc" are
given by the best fit values  WMAP3 $+$ SDSS, the underlined "S"
is by WMAP3 $+$ SDSS $+$ SNLS and "R" by WMAP3 $+$ SDSS $+$ Riess
sample. The right panels show the corresponding cases where within
the 2$\sigma$ allowed regions the oscillating effects are
relatively eminent. Same as the above conventions in the upper
panels the blue lines dubbed "SNAP" illustrate the detectability
of future SNAP\cite{snap}. It is noteworthy that the combination
with SNLS gives relatively the most stringent constraints. We can
find that for the case with WMAP $+$ SDSS only, a larger
oscillation is allowed and there are some oscillating features on
the (residue) Hubble diagram. And the effects of the "Most" case
in Table 1 are more eminent, which could be possibly detected even
by the CURRENTLY ONGOING SNIa observations. In the cases combined
with Riess "gold" sample or with SNLS, the oscillating effects are
less eminent. Nevertheless, oscillations are still present and in
large areas of the parameter spaces, SNAP will be able to detect
such features\footnote{In some sense in Table 1 for the "Most"
case by WMAP $+$ SDSS combination only, the relevant parameter
space will get relatively stringently constrained by combinations
with SNLS or with Riess "gold" sample. However given the possible
discrepancies between SNLS and the Riess sample, which has been
somewhat illustrated in Ref. \cite{Spergel:2006hy}, currently
ongoing SNIa observations can in some sense detect the
(semi)-oscillating features in dark energy EOS as allowed by WMAP
$+$ SDSS combination. }.

Now we turn to the studies on the local bump-like features of dark
energy EOS. It is physically intuitive that dark energy EOS might
not be exactly periodic, instead it could be semi-periodic or even
with features on some specific redshifts. The parametrization
adopted in Eq. (\ref{Wqbump}) ( $w= w_0 + A (\ln a - \lambda)^3
\exp(- (\ln a - \lambda)^4/d)$ ) can accomodate constant EOS in
cases where $A=0$ . $\lambda$ determines the locations of bumps,
$d$ determines the width and $A$ determines the amplitudes of the
bumps. One can easily understand that at the point $\ln a =
\lambda$ we have $w=w_0$ and for $\ln a < \lambda$ (high
redshifts) there will be a trough and for $\ln a
> \lambda$ (low redshifts) there will be a peak (this happens only
in cases where $\lambda<0$, which is the region of our interest).
And on very high redshifts where $\ln a \ll \lambda$  the second
term will get damped exponentially and w approaches the value of
$w_0$. For simplicity and in the illustrative study here we fix
$w_0=-1$.

\begin{figure}[htbp]
\includegraphics[scale=0.5]{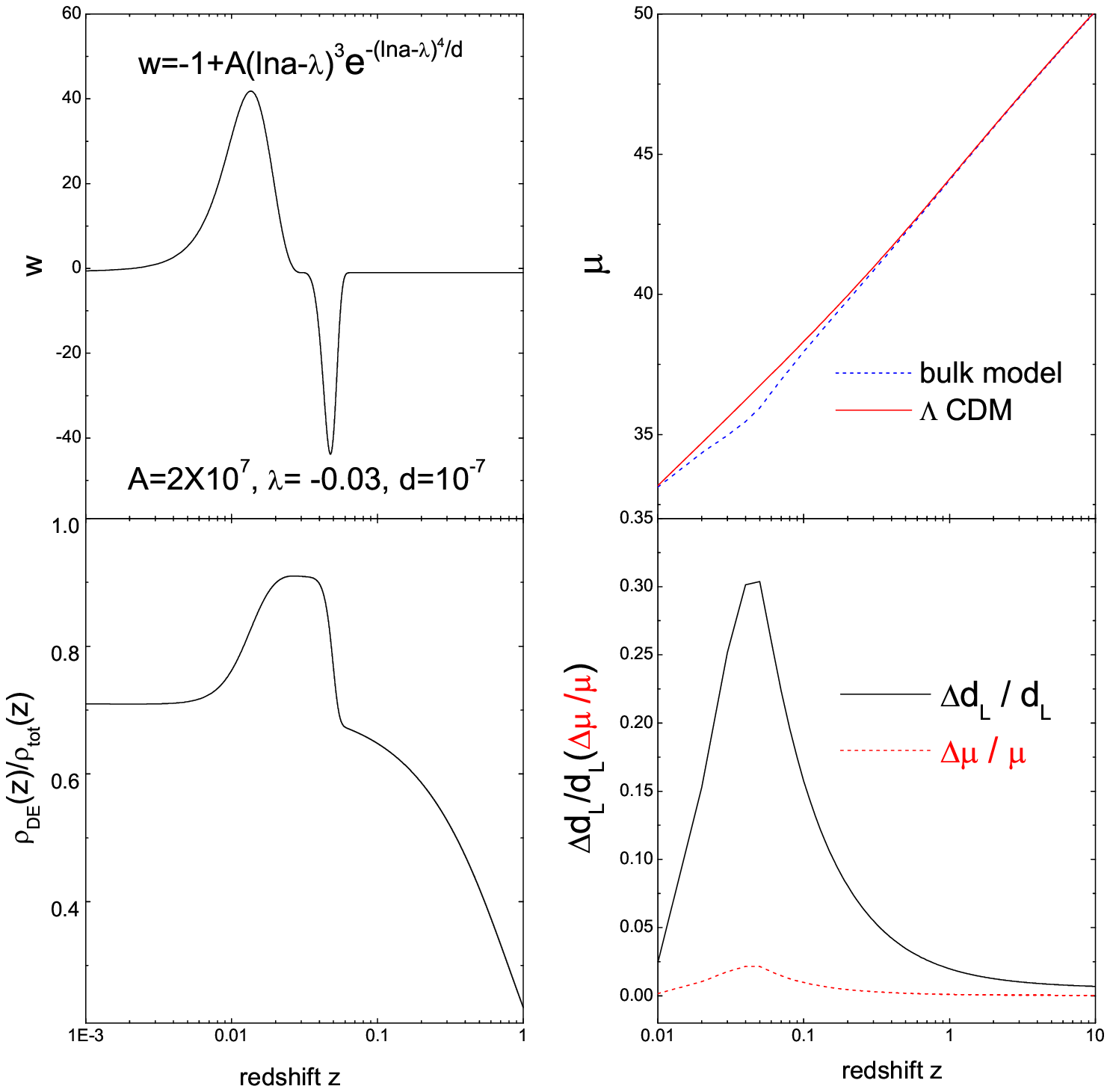}
\includegraphics[scale=0.55]{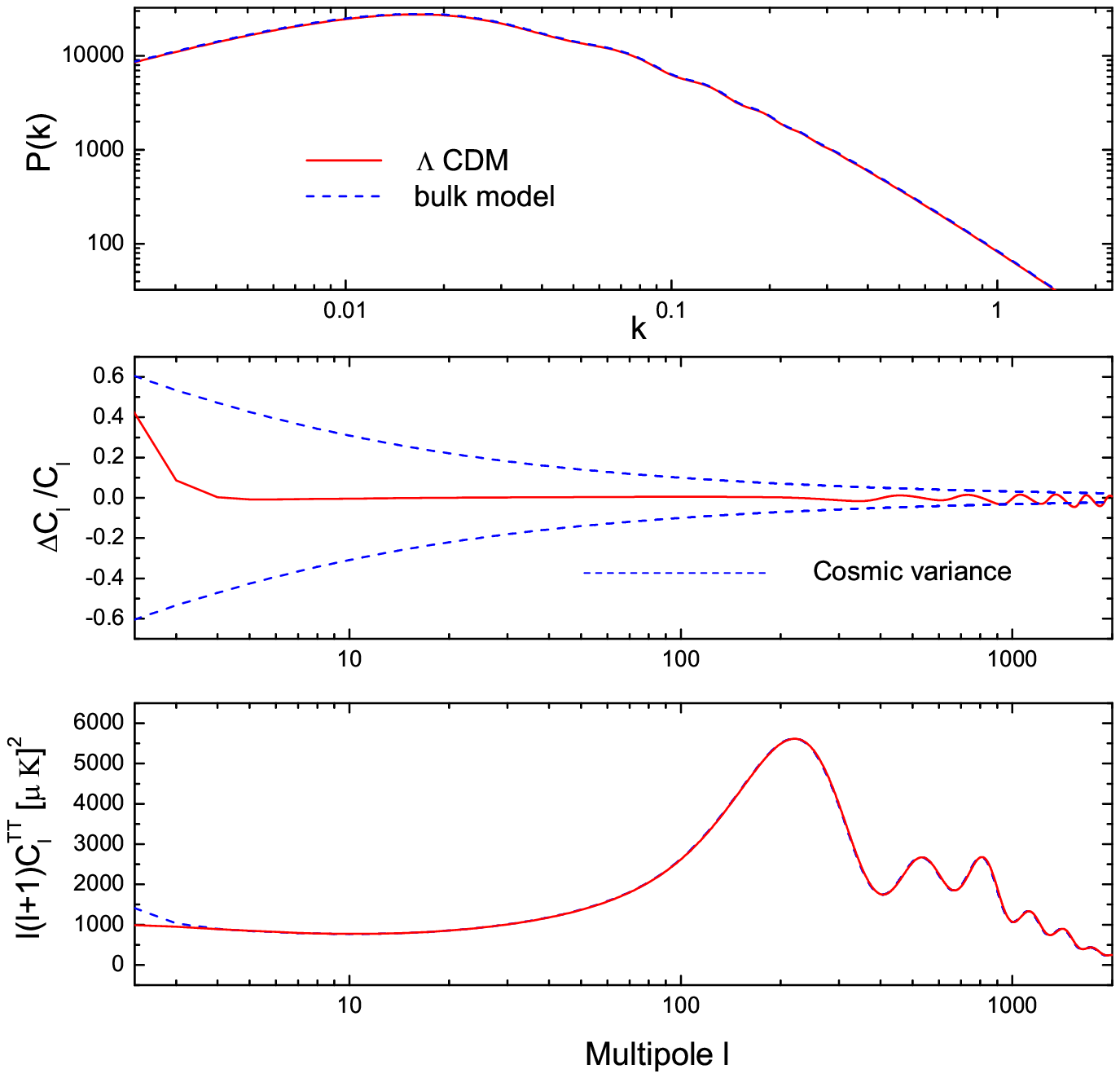}
\caption{ Imprints of bump-like dark energy EOS (denoted by "bulk
model") on the (residual) Hubble diagram, dark energy density
fractions (left) and the corresponding effects on CMB and LSS
(right), shown together with the imprints by the $\Lambda$CDM
model. The bump takes place at low redshifts which can be
detectable by (future) SNIa observations. The effects on CMB are
indistinguishable from the corresponding $\Lambda$CDM model beyond
cosmic variance, and nor can LSS tell one (bump-like) from the
other ($\Lambda$CDM). \label{fig:fig1}}
\end{figure}

Firstly we consider the cases where the bump takes place at low
redshifts which can be detectable by (future) SNIa observations.
In Fig.\ref{fig:fig1} we delineate the imprints of bump-like dark
energy EOS on the (residual) Hubble diagram, dark energy density
fractions (left) and the corresponding effects on CMB and LSS
(right). We have chosen $A=2\times 10^7, \lambda=-0.03$ and
$d=10^{-7}$ for the first illustration. From Fig.\ref{fig:fig1} we
can find that for such a specific choice of parameters the effects
on CMB are indistinguishable from the corresponding $\Lambda$CDM
model beyond cosmic variance, and nor can LSS tell one (bump-like)
from the other ($\Lambda$CDM). It is understandable that in such a
case as the effects of the peak and trough somewhat cancel out in
the contributions to $w_{eff}$ defined in Eq. (\ref{weff}), the
effects on CMB and LSS would then be {\it almost}
indistinguishable from the $\Lambda$CDM case, and the results are
consistent with those in Refs.\cite{Wang:1999fa,Zhao:2005vj}. In
such a case, observations with geometric constraints on the
redshift "tomography" will be of great importance to break such a
degeneracy. The resulting (residual) Hubble diagrams are different
from the $\Lambda$CDM model, and observations like SNAP can
possibly detect such features.

\begin{figure}[htbp]
\includegraphics[scale=1]{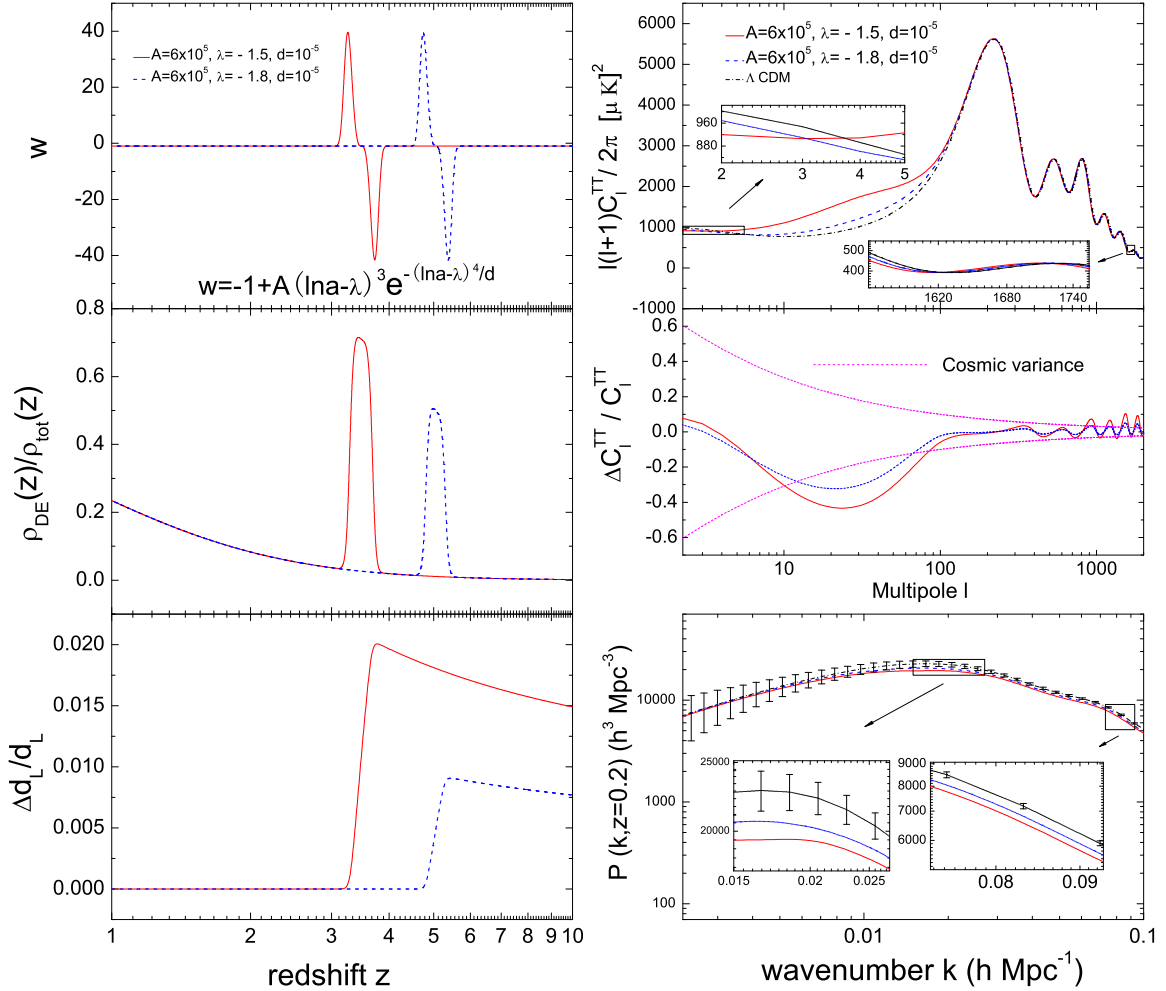}
\caption{ Imprints of bump-like dark energy EOS on  CMB, LSS (right
panels), on dark energy density fractions and on the residual Hubble
diagram (left panels). The bumps take place at high redshifts which
cannot be detectable by (future) SNIa observations. On the other
hand, cosmic variance limited CMB (e.g. WMAP3) and
LAMOST\cite{lamost} are possible to detect such features. Note here
we have not performed a global analysis and did not take into
account the parameter degeneracies \label{fig:fig3} .}
\end{figure}

Although SNIa in some sense makes the only direct detection of dark
energy, it is believed that due to some unavoidable effects like by
dust the redshifts probed by SNIa cannot be too high, and for
example in SNAP simulations one typically takes $z\le 1.7$. In cases
when the bump-like features take place at higher redshifts, we
cannot expect to probe such features with even future SNAP
observations. On the other hand as the current observations are
consistent with a $\Lambda$-like dark energy, typically we cannot
hope dark energy will be non-negligible in epochs far before the
matter-radiation equality epoch, although such a possibility remains
(for relevant studies see e.g. \cite{Malquarti:2002bh,Wang:2003yj}).
On the other hand with our parametrized EOS given in Eq.
(\ref{Wqbump}) we cannot expect dark energy component to be
significant on very high redshifts, especially for cases where
$w_0=-1$\footnote{This is however, still possible given some fine
tunings on the remaining parameters in Eq. (\ref{Wqbump}), which
will not be explored by the current paper. }. Under such a
circumstance we take the locations of bumps to be slightly larger
than $z=2$: in one case $\lambda=-1.5$ and in another case
$\lambda=-1.8$. These correspond to $z\simeq 3.5$ and 5.0
respectively where $\ln a = \lambda$. For both of the cases we have
fixed $A=6\times 10^5$ and $d=10^{-5}$. In Fig.\ref{fig:fig3} we
show the imprints of bump-like dark energy EOS on CMB, LSS (right
panels), on dark energy density fractions and on the residual Hubble
diagram (left panels). The bumps take place at high redshifts which
cannot be detectable by (future) SNIa observations. On the other
hand, observations of Gamma-ray bursts (GRB) can probe relatively
higher redshifts than
SNIa\cite{Takahashi:2003ap,Dai:2004tq,Hooper:2005xx}, with the
accumulations of GRB events as by SWIFT and better understandings on
the systematics one is hopefully able to detect such features
through geometric observations on the Hubble diagrams. Moreover
observations like the 21 cm tomography
\cite{Field59,Loeb:2003ya,Chen:2003gc,Pen:2004de} are potentially
able to give almost a large number of independent samples compared
with CMB and LSS observations\cite{Loeb:2003ya}, features of
bump-like dark energy EOS on intermediate redshifts are promisingly
detected by observations of the 21 cm tomography. In the right
panels of Fig.\ref{fig:fig3} we find dramatically that bump-like
features are possibly detectable by cosmic variance limited CMB
(e.g.
WMAP3\cite{Spergel:2006hy,Page:2006hz,Hinshaw:2006,Jarosik:2006,WMAP3IE})
and LAMOST\cite{lamost}.  Part of the reason lies on the fact that
with a different width and a higher redshift compared with the
previous example in Fig.\ref{fig:fig1}, here the energy density
fraction $\Omega_{DE}(a)$ increases and decreases rapidly and the
magnitude of the peak value of $\Omega_{DE}(a)$ is somewhat lower,
the asymmetric shapes of the peak and trough cannot compensate
thoroughly to give $w_{eff}$ close to $-1$. In fact one can find for
our parametrization given in Eq. (\ref{Wqbump}) one will typically
get $w_{eff}>-1$ and in cases with larger $-\lambda$ the deviation
of $w_{eff}$ from -1 can be larger, which will also lead to some
shifts on CMB peaks, as shown in Fig.\ref{fig:fig3}. It is
noteworthy to point out that both the first year WMAP and WMAP3 show
some local glitches out of cosmic variance at $l\sim 20-40$. In our
case of Fig.\ref{fig:fig3} there are also some relevant features on
such scales and this deserves  further investigations. While such
features can be affected by different foreground analysis and so
forth\cite{Slosar:2004fr,deOliveira-Costa:2006zj,Eriksen:2006xr}, it
is theoretically possible that the features of WMAP TT on scales
$l\sim 20-40$ are due to some bump-like features of dark energy EOS,
or some semi-oscillations. Note here we have fixed the background
parameters rather than performing a global analysis, and in cases we
take into account the parameter degeneracies the detectability of
the observations would typically get weaker. On the other hand we
can expect that in the global analysis when all of the parameters
(including the bias factor) can vary, the observations can give
constraints or detect the signatures where the amplitudes of bumps
are even higher than the examples listed above.

\section{discussion and conclusion}

We should point that in all our three data combinations, a large
area of oscillating Quintom where the EOS of dark energy gets across
$-1$ during time evolutions are allowed by the current observations,
which is different from the previous case in Ref.\cite{DSPRL} and
from the case in Ref. \cite{Barenboim:2004kz}. Moreover contrary to
Refs.\cite{DSPRL,Barenboim:2004kz}, oscillating Quintom displays the
distinctive feature that due to the average effects on the regions
of EOS where $w>-1$ and $w<-1$, the quantity $w_{eff}$ can be very
close to $-1$. Oscillating Quintom-like dark energy, where the
unification of two epochs of accelerated expansions can be realized
and the coincidence problem can be solved\cite{Feng:2004ff}, proves
to be a good fit to current cosmology. We should stress again that
resembling the oscillating primordial spectrum
case\cite{Okamoto:2003wk}, the not good enough convergence of
oscillating dark energy EOS is inevitable in the light of the
current observations. In the WMAP $+$ SDSS allowed parameter space
there are areas with significant oscillations, which implies that
the CURRENTLY ONGOING SNIa observations may detect such oscillating
or semi-oscillating features on the (residual) Hubble diagram.
Moreover while oscillating features on CMB might be explained by
both oscillating primordial spectrum and oscillations in dark energy
EOS, the oscillating features on the Hubble diagram cannot be due to
features in the primordial spectrum, but by oscillating EOS.  Local
bumps of dark energy EOS may leave some distinctive imprints on CMB,
LSS and SNIa. The bumps are potentially detectable by geometrical
observations like SNIa and GRB. Future observations of 21 cm
tomography open a very promising window to detect/exclude such
features. LSS measurements like LAMOST and cosmic variance limited
CMB  may also detect such features. In particular, bump-like dark
energy EOS on high redshifts {\it might} be responsible for the
features of WMAP on ranges $l \sim 20-40$, which is interesting and
deserves addressing further.

Given the fact that currently we know relatively very little on
theoretical aspects of dark energy, observing dynamical dark
energy is currently the most important aspect of dark energy
study. On the observational probes typically one needs some
parametrizations on the form of dark energy EOS. While using some
specific forms of parametrizations one often takes the risks of
getting biased ( see e.g. \cite{Bassett:2004wz} ), investigations
towards unbiased probe of
DE\cite{Huterer:2002hy,Wang:2005ya,Simon:2004tf} in light of all
the available cosmological observations still need further
developments. On the other hand some of the parametrizations are
in some sense well motivated (see e.g. \cite{Bassett:2004wz}), and
in fittings starting with parametrized EOS one typically has
larger $\nu$ (number of data minus the number of parameters) and
hence gets better constraints on dark energy. With the
accumulations of observational data poor parametrizations will get
ruled out and in this sense parametrized study of dark energy
provides a complementary study of the non-parametric probes. As
from any quintessence-like or phantom-like EOS which do not get
across $-1$ one can reconstruct the dark energy
potential\cite{Lewisnotes}, any parametrizations of
non-Quintom-like EOS correspond to some specific forms of
quintessence/phantom potentials\footnote{In cases where $w>1$ they
correspond to quintessence with negative potentials.} and hence
such parametrizations are somewhat well motivated. Bump-like EOS
which do not get across $-1$ also correspond to some specific DE
potentials which can be straightforwardly worked out. Similarly
for (semi-)oscillating EOS which do not get across $-1$ one can
work out the corresponding DE potentials. In cases with bumps
described in Eq. (\ref{Wqbump}) and with oscillations in Eq.
(\ref{Wqosc}), one cannot simply reconstruct the potentials of DE
due to the distinctive nature of
Quintom\cite{Feng:2004ad,Zhao:2005vj}. However this remains
possible for example the models with high derivatives
\cite{Li:2005fm,Zhang:2006ck} and in the framework with modified
gravity\cite{MorikawaOSC,Dvali:2000hr,Sahni:2002dx,Perivolaropoulos:2003we,Stefancic:2003bj,Perivolaropoulos:2005yv}.


 {\bf{Acknowledgments:}} We acknowledge the use of the Legacy
Archive for Microwave Background Data Analysis (LAMBDA). Support for
LAMBDA is provided by the NASA Office of Space Science. We have
performed our numerical analysis on the Shanghai Supercomputer
Center (SSC). We used a modified version of
CAMB\cite{Lewis:1999bs,IEcamb} which is based on
CMBFAST\cite{cmbfast,IEcmbfast}. We are grateful to Hiranya Peiris,
Yunsong Piao and Lifan Wang for discussions related to this project.
We thank Xuelei Chen, Yaoquan Chu and Long-long Feng for discussions
on LAMOST. We thank Sarah Bridle, Peihong Gu, Steen Hannestad,
Antony Lewis, Mingzhe Li, Yongzhong Xu, Jun'ichi Yokoyama, Max
Tegmark and Penjie Zhang for helpful discussions. B. F. would like
to thank the hospitalities of IHEP during his visit to Beijing. This
work is supported in part by National Natural Science Foundation of
China under Grant Nos. 90303004, 10533010 and 19925523 and by
Ministry of Science and Technology of China under Grant No. NKBRSF
G19990754. B. F. is supported by the JSPS fellowship program.

\newcommand\AJ[3]{~Astron. J.{\bf ~#1}, #2~(#3)}
\newcommand\APJ[3]{~Astrophys. J.{\bf ~#1}, #2~ (#3)}
\newcommand\APJL[3]{~Astrophys. J. Lett. {\bf ~#1}, L#2~(#3)}
\newcommand\APP[3]{~Astropart. Phys. {\bf ~#1}, #2~(#3)}
\newcommand\CQG[3]{~Class. Quant. Grav.{\bf ~#1}, #2~(#3)}
\newcommand\JETPL[3]{~JETP. Lett.{\bf ~#1}, #2~(#3)}
\newcommand\MNRAS[3]{~Mon. Not. R. Astron. Soc.{\bf ~#1}, #2~(#3)}
\newcommand\MPLA[3]{~Mod. Phys. Lett. A{\bf ~#1}, #2~(#3)}
\newcommand\NAT[3]{~Nature{\bf ~#1}, #2~(#3)}
\newcommand\NPB[3]{~Nucl. Phys. B{\bf ~#1}, #2~(#3)}
\newcommand\PLB[3]{~Phys. Lett. B{\bf ~#1}, #2~(#3)}
\newcommand\PR[3]{~Phys. Rev.{\bf ~#1}, #2~(#3)}
\newcommand\PRL[3]{~Phys. Rev. Lett.{\bf ~#1}, #2~(#3)}
\newcommand\PRD[3]{~Phys. Rev. D{\bf ~#1}, #2~(#3)}
\newcommand\PROG[3]{~Prog. Theor. Phys.{\bf ~#1}, #2~(#3)}
\newcommand\PRPT[3]{~Phys.Rept.{\bf ~#1}, #2~(#3)}
\newcommand\RMP[3]{~Rev. Mod. Phys.{\bf ~#1}, #2~(#3)}
\newcommand\SCI[3]{~Science{\bf ~#1}, #2~(#3)}
\newcommand\SAL[3]{~Sov. Astron. Lett{\bf ~#1}, #2~(#3)}

{}

\end{document}